\documentclass[aps,showpacs,preprintnumbers,amsmath, amssymb]{revtex4}

\usepackage{float}
\usepackage{graphics,epsfig}
\usepackage{graphicx}
\usepackage{dcolumn}
\usepackage{bm}

\begin{document}
\baselineskip=0.8 cm
\title{{\bf Refractive index in generalized superconductors with Born-Infeld electrodynamics}}

\author{Jun Cheng, Qiyuan Pan\footnote{panqiyuan@126.com}, Hongwei Yu\footnote{Corresponding author at hwyu@hunnu.edu.cn}, Jiliang Jing\footnote{jljing@hunnu.edu.cn}} \affiliation{Key
Laboratory of Low Dimensional Quantum Structures and Quantum Control
of Ministry of Education, Synergetic Innovation Center for Quantum
Effects and Applications, and Department of Physics, Hunan Normal
University, Changsha, Hunan 410081, China}

\vspace*{0.2cm}
\begin{abstract}
\baselineskip=0.6 cm
\begin{center}
{\bf Abstract}
\end{center}

We  investigate,  in the
probe limit,  the negative refraction in the generalized superconductors with the Born-Infeld
electrodynamics.
We observe that the system has a negative
Depine-Lakhtakia index in the superconducting phase at small
frequencies and the greater the Born-Infeld corrections  the larger the range of
frequencies or the range of temperatures  for which the negative
refraction occurs. Furthermore, we find that the tunable
Born-Infeld parameter can be used to improve the propagation of light in the
holographic setup. Our analysis indicates that the Born-Infeld
electrodynamics plays an important role in determining the
optical properties of the boundary theory.

\end{abstract}
\maketitle
\newpage
\vspace*{0.2cm}

\section{Introduction}

As one of the most significant developments in the theoretical
physics over the last decade, the anti-de Sitter/conformal field
theories (AdS/CFT) correspondence
\cite{Maldacena,Witten,Gubser1998}, which provides an astonishing
duality between the gravity in the $d$-dimensional spacetime and the
gauge field theory living on its ($d-1$)-dimensional boundary, has
been employed to study the strong coupled systems in condensed
matter physics which are intractable by the traditional approaches
\cite{JZaanen}. In particular, such a strong/weak duality might give
some insights into the pairing mechanism in the high $T_{c}$
superconductors \cite{HartnollJHEP12}. It was first suggested by
Gubser~\cite{GubserPRD78} that the spontaneous $U(1)$ symmetry
breaking by bulk black holes can be used to construct gravitational
duals of the transition from a normal state to a superconducting state
in the boundary theory, which exhibit characteristic  behaviors of
superconductors \cite{HartnollPRL101}. Along this line, there have
been a lot of works studying various gravity models with the
property of the so-called holographic superconductor, for reviews,
see Refs. \cite{HartnollRev,HerzogRev,HorowitzRev,CaiRev} and
references therein.

On the other hand, the refractive index, which reflects the
propagation property of light in an electromagnetic medium, is one
of the most important electromagnetic properties of a medium. In
1968, Veselago first considered the case of a medium which has both
negative dielectric permittivity and negative magnetic permeability
at a given frequency and proposed in theory that the refraction
index might be negative \cite{Veselago}. Around 2000, this exotic
electromagnetic phenomenon---negative refraction, which implies that
the energy flux of the electromagnetic wave flows in the opposite
direction with respect to the phase velocity, was experimentally
realized in a new class of artificial media commonly called
``metamaterials" \cite{Smith,Pendry}.  Since then, the study on the
negative refraction has attracted intensive interest and a large
number of the negative refractive index materials have been
introduced \cite{Ramakrishna}. Interestingly, recent studies show
that in the radio, microwave and low-terahertz frequency ranges the
superconductors can behave as metamaterials and exhibit negative
refraction \cite{Anlage}. Using the AdS/CFT correspondence, Gao and
Zhang investigated the optical properties in the s-wave
superconductors and found that the negative refraction does not
appear in these holographic models in the probe limit
\cite{GaoZhang}. However, in the fully backreacted spacetime,
Amariti \emph{et al.} showed that the negative refraction is allowed
in the superconducting phase \cite{AmaritiFMS2011}. Extending the
investigation to the generalized holographic superconductors in
which the spontaneous breaking of a global $U(1)$ symmetry occurs
via the St\"{u}ckelberg mechanism \cite{Franco,FrancoJHEP},
Mahapatra \emph{et al.} observed that a negative Depine-Lakhtakia
index may appear at low frequencies in the theory dual to the
R-charged black hole, which indicates that the system exhibits
negative refraction even in the probe limit
\cite{MahapatraJHEP2014}. Other investigations based on the
refractive index in the holographic dual models can be found, for
example, in Refs.
\cite{AmaritiFMP2011,GJSJHEP2011,LiuPRD2011,AmaritiFMP2013,JiangPRD2013,PhukonSJHEP2013,DeyMT2014,ForcellaJHEP2014,MahapatraJHEP2015,JiangPRD2016}.

The aforementioned works on the refractive index in the holographic
superconductor models are based on the usual Maxwell
electrodynamics. Considering  the Maxwell theory as only a special
case of or a leading order approximation to  nonlinear
electrodynamics, we may extend the investigation to the nonlinear
electrodynamics which essentially implies  higher derivative
corrections of the gauge field \cite{HendiJHEP}. As a first attempt
to investigate how the nonlinear electrodynamics affects the
properties of the holographic dual model, Jing and Chen introduced
the holographic superconductors in the Born-Infeld electrodynamics
and observed that the nonlinear Born-Infeld corrections  make it
harder for the scalar condensation to form \cite{JS2010}. Later the
holographic dual models with the Power-Maxwell electrodynamics
\cite{JingJHEP}, the Maxwell field strength corrections \cite{PJWPRD},
the logarithmic form of nonlinear electrodynamics \cite{JPCPLB} and
the exponential form of nonlinear electrodynamics \cite{ZPCJNPB} were
realized and some interesting properties were disclosed. Considering
the increasing interest in the study of the holographic dual models
with the nonlinear electrodynamics
\cite{JingPRD2011,SDSL2012JHEP156,LeeEPJC,LPW2012,Roychowdhury,BGQX,BGRLPRD2013,YaoJing,
DLAP,SGMPLA,Chaturvedi2015,LaiPLB2015,LGW2016,Ghorai2016,JingCQG2016,SheykhiShaker2016,WuZhang20162017,Sherkatghanad,Gangopadhyay2017},
in this paper, we are going to examine the influence of the
Born-Infeld electrodynamics on the optical properties of the
generalized holographic superconductors in the probe limit. We will
show that the Born-Infeld corrections affect not only the range of
frequencies or the range of temperatures for which negative
refraction occurs but also the dissipation  in the system,
and this is helpful for us to understand the influences of the $1/N$
($N$ is the color quantum number) or $1/\lambda$ ($\lambda$ is the
't Hooft coupling) corrections on the holographic superconductor
models and their optical properties.

The structure of this work is as follows. In Sec. II we will
construct the generalized holographic superconductors with
the Born-Infeld electrodynamics in the probe limit and analyze the
effect of the Born-Infeld electrodynamics on the condensate of the
system. In Sec. III we will consider the effect of the Born-Infeld
electrodynamics on the negative refraction in generalized
superconductors. We will summarize our results in the last section.

\section{Generalized holographic superconductors with Born-Infeld electrodynamics}

In order to construct the generalized holographic superconductors
with the Born-Infeld electrodynamics in the probe limit,  we consider a four-dimensional planar
Schwarzschild-AdS black hole background
\begin{eqnarray}\label{SchBH}
ds^2=-f\left(r\right)dt^2+\frac{1}{f\left(r\right)}dr^2+r^2\left(dx^2+dy^2\right),
\end{eqnarray}
with
\begin{eqnarray}
f(r)=r^2\left(1-\frac{r_{h}^{3}}{r^{3}}\right),
\end{eqnarray}
where $r_{h}$ is the radius of the event horizon. The Hawking
temperature of the black hole, which will be interpreted as the
temperature of the CFT, is determined by $T=3r_{h}/4\pi$.
 We then introduce the Born-Infeld
electrodynamics and a charged scalar field coupled via a
generalized Lagrangian
\begin{equation}
S=\int d^{4}x\sqrt{-g}\left[\frac{1}{b}\left(1-\sqrt{1+\frac{b
F^2}{2}}\right)-\frac{(\partial_{\mu}\Psi)^2}{2}-\frac{m^2\Psi^2}{2}-|\textrm{G}(\Psi)|(\partial\alpha-A)^2\right],
\label{Lagrangian}
\end{equation}
where both the charged scalar field $\Psi$ and the phase $\alpha$
are real, and the local $U(1)$ gauge symmetry in this theory is
given by $A_{\mu}\rightarrow A_{\mu}+\partial_{\mu}\lambda$ together
with $\alpha\rightarrow \alpha+\lambda$. When the Born-Infeld
parameter $b\rightarrow0$, the model (\ref{Lagrangian}) reduces to
the generalized holographic superconductors with the usual Maxwell
electrodynamics investigated in
\cite{Franco,FrancoJHEP,MahapatraJHEP2014}.

Using the gauge symmetry to fix the phase $\alpha=0$ and taking the
ansatz $\Psi=\Psi(r)$, $A=\Phi(r)dt$, we can get the equations of
motion
\begin{align}
\Psi''+\biggl(\frac{2}{r}+\frac{f'}{f}\biggr)\Psi'-\frac{m^2}{f}\Psi+\frac{\Phi^2}{f^{2}}\frac{d\textrm{G}(\Psi)}{d\Psi}=0,
\label{PsiEquationMotion}
\end{align}
\begin{align}
\Phi''+\frac{2}{r}(1-b\Phi'^2)\Phi'-\frac{2\textrm{G}(\Psi)}{f}(1-b\Phi'^2)^{3/2}\Phi=0,
\label{PhiEquationMotion}
\end{align}
where a prime denotes the derivative with respect to $r$. Imposing
the appropriate boundary conditions, we can solve Eqs.
(\ref{PsiEquationMotion}) and (\ref{PhiEquationMotion}) numerically
by doing integration from the horizon out to the infinity
\cite{HartnollJHEP12}. At the horizon $r=r_{h}$, we require the
regularity conditions
\begin{equation}
\Phi(r_{h})=0,~~~~\Psi'(r_{h})=\frac{m^2\Psi(r_{h})}{f'(r_{h})}.
\end{equation}
At infinity $r\rightarrow\infty$, we have asymptotic behaviors
\begin{equation}
\Phi=\mu-\frac{\rho}{r},~~~~\Psi=\frac{\Psi_{-}}{r^{\Delta_{-}}} +
\frac{\Psi_{+}}{r^{\Delta_{+}}},
\end{equation}
with $\Delta_\pm=(3\pm\sqrt{9+4m^2})/2$. According to the AdS/CFT
correspondence, $\mu$ and $\rho$ are interpreted as the chemical
potential and the charge density in the dual field theory respectively.
Note that, provided $\Delta_{-}$ is larger than the unitarity bound,
both $\Psi_{-}$ and $\Psi_{+}$ can be normalizable and  be
used to define operators on the dual field theory, $\Psi_{-}=\langle
O_{-}\rangle$, $\Psi_{+}=\langle O_{+}\rangle$, respectively
\cite{HartnollJHEP12}. In this work, we impose the boundary condition
$\Psi_{-}=0$ since we concentrate on the condensate for the operator
$\langle O_{+}\rangle$. Considering that the choices of the scalar
field mass will not qualitatively modify our results, we will set
$m^2=-2$ for concreteness. Thus, the scalar condensate is now
described by the operator $\langle O_{2}\rangle=\Psi_{2}$.

\begin{figure}[ht]
\includegraphics[scale=0.6]{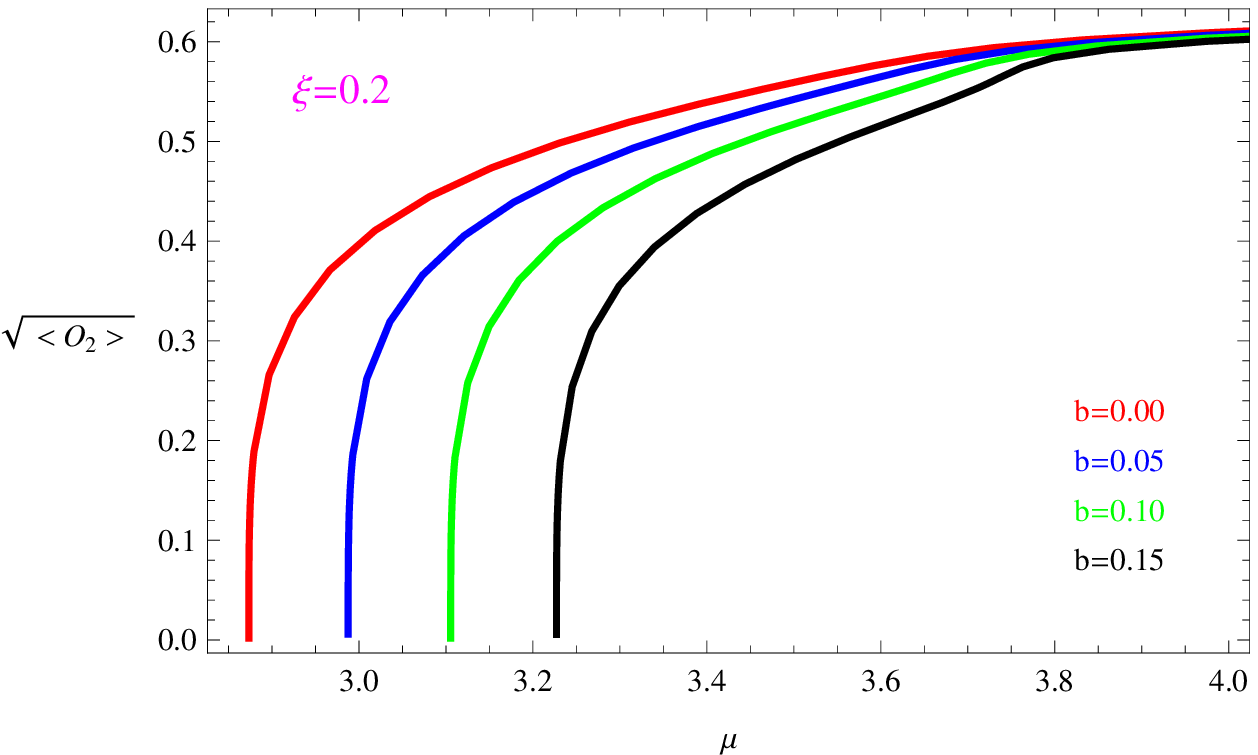}\hspace{0.2cm}%
\includegraphics[scale=0.6]{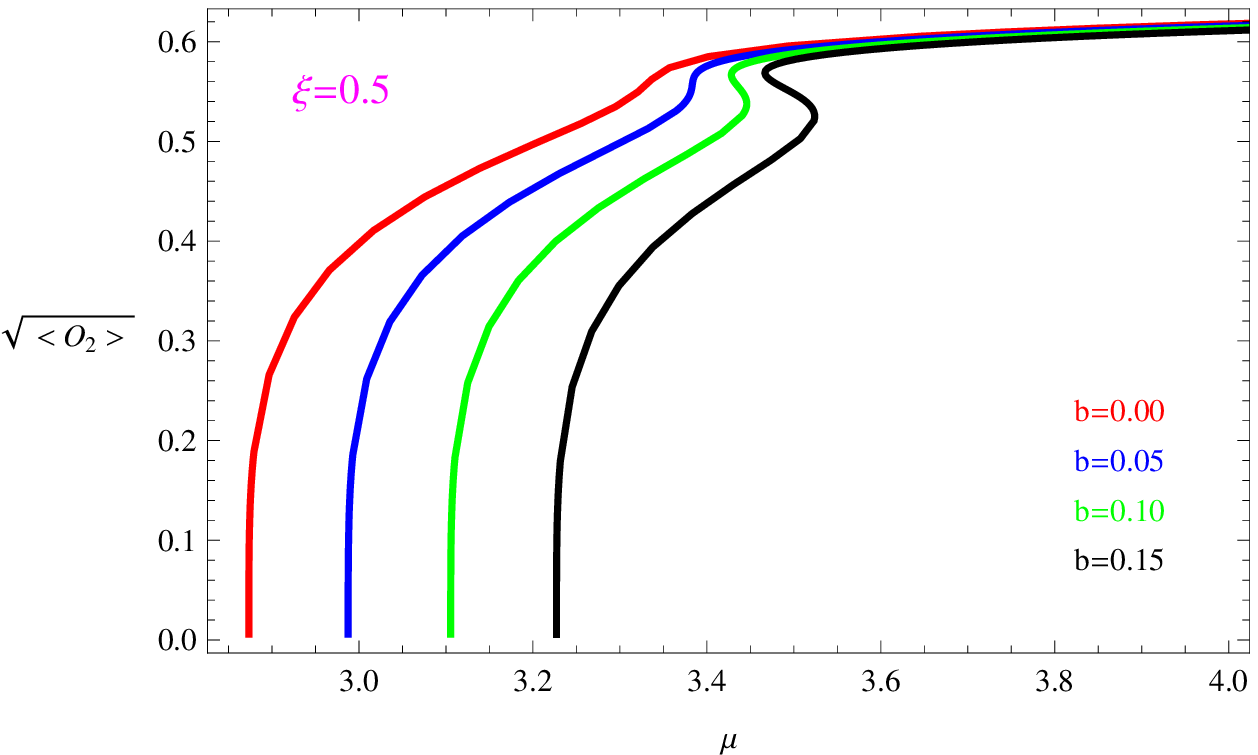}\\ \vspace{0.0cm}
\caption{\label{Condensate} (Color online) The condensate as a
function of the chemical potential with the fixed coefficient
$\xi=0.2$ (left) and $\xi=0.5$ (right) for different values of the
Born-Infeld parameter $b$. The four lines in each panel from left to
right correspond to increasing Born-Infeld parameter, i.e., $b=0.00$
(red), $0.05$ (blue), $0.10$ (green) and $0.15$ (black)
respectively. For concreteness, we have set the mass of the scalar
field $m^{2}=-2$.}
\end{figure}

We will investigate the effect of the Born-Infeld electrodynamics on
the condensate of the system. For simplicity, we consider a
particular form of $\textrm{G}(\Psi)$, i.e.,
\begin{equation}
\textrm{G}(\Psi)=\Psi^{2}+\xi\Psi^{8},
\end{equation}
with the model parameter $\xi$ introduced in Ref.
\cite{MahapatraJHEP2014}. 
  We solve the equations of motion
numerically, and plot in Fig. \ref{Condensate} the condensate around
the critical region for chosen values of $\xi$ and various
Born-Infeld parameters. In the case of $\xi=0$ or small $\xi$, for
example $\xi=0.2$, the transition is of the second order and the
condensate approaches zero as $\langle O_{2}\rangle\sim
(\mu-\mu_{c})^{1/2}$ for all values of $b$ considered here. However,
the story is different if we increase the model parameter $\xi$.
Focusing on the case of $\xi=0.5$ in Fig. \ref{Condensate}, as an
example, we observe that $\langle O_{2}\rangle$ becomes multivalued
near the critical point and the first-order phase transition
appears. This behavior keeps for all values of $b$ except for
$b=0.00$ and increasing $b$ makes the behavior  more  distinct,
which supports the findings in Ref. \cite{JingPRD2011} and indicates
that the greater the  Born-Infeld corrections the  easier it is  for
the first-order phase transition to emerge. Thus, we find that not
only the model parameter $\xi$ but also the Born-Infeld parameter
$b$ can manipulate the order of the phase transition. The
Born-Infeld electrodynamics provides richer physics in terms of  the
phase transition. On the other hand, we can see clearly from Fig.
\ref{Condensate} that the critical chemical potential $\mu_{c}$
increases with the increase of the Born-Infeld parameter for a fixed
model parameter $\xi$, which indicates that  large Born-Infeld
electrodynamics corrections  hinder the formation of the
condensation. This agrees well with the findings in the first
holographic superconductor model with the Born-Infeld
electrodynamics introduced in \cite{JS2010}.

\section{Negative refraction in generalized superconductors with Born-Infeld electrodynamics}

In the preceding  section we have constructed the generalized
holographic superconductors with the Born-Infeld electrodynamics.
Now we are in a position to discuss the optical properties of these
holographic systems and reveal the effect of the Born-Infeld
electrodynamics on the negative refraction in generalized
superconductors. Just as the standard assumption in the usual
Maxwell electrodynamics \cite{AmaritiFMP2011,AmaritiFMP2013}, we
will assume that the boundary theory is weakly coupled to a
dynamical electromagnetic field since the Born-Infeld
electrodynamics is a correction to the usual Maxwell  electrodynamics  and
calculate the refractive index of the system perturbatively.

\subsection{Holographic setup}

Using the linear response theory (for more details see Ref.
\cite{AmaritiFMP2011}), we can find that the electric permittivity
and the effective magnetic permeability for  isotropic media are
determined by the frequency dependent transverse current correlators
as follows
\begin{eqnarray}
&&\epsilon \left( \omega \right) = 1+\frac{4\pi }{\omega ^{2}}
C_{em}^{2}G_{T}^{0}\left( \omega \right),
\nonumber \\
&&\mu\left( \omega \right) = \frac{1}{1- 4\pi
C_{em}^{2}G_{T}^{2}\left( \omega \right) }, \label{epsmu}
\end{eqnarray}
where $C_{em}$ is the electromagnetic coupling which will be set to
unity when performing numerical calculations, and $G_{T}^{0}\left(
\omega \right)$ and $G_{T}^{2}\left( \omega \right)$ are the
expansion coefficients of the retarded correlators
\cite{SonStarinets} in the spatial momentum $k$, i.e.,
\begin{eqnarray}
G_{T}\left( \omega,k \right) =G_{T}^{0}\left( \omega \right) +k^{2}
G_{T}^{2}\left( \omega \right) + \cdots. \label{GT}
\end{eqnarray}
So the refractive index can be given by
\begin{eqnarray}
n^{2}(\omega)=\epsilon(\omega)\mu\left(\omega\right).
\label{RefractiveIndex}
\end{eqnarray}
Generally, the existence of  a negative refractive index can be
predicted by using the Depine-Lakhtakia (DL) index $n_{DL}$
expressed as \cite{DepineLakhtakia}
\begin{eqnarray}\label{DLIndex}
n _{DL}= Re[\epsilon(\omega)]|\mu(\omega)| +Re[\mu(\omega)]|
\epsilon(\omega)|,
\end{eqnarray}
where the negativity of the DL index indicates that the phase
velocity in the medium is opposite to the direction of the energy flow,
i.e., the system has negative refractive index.

Let us now move on to the strategy to calculate these quantities in our
holographic superconducting system with the Born-Infeld electrodynamics.
Considering the gauge field perturbation
\begin{eqnarray}
\delta A_{x}= A_{x}(r) e^{-i \omega t + iky},
\end{eqnarray}
we obtain the equation of motion for $A_{x}(r)$ as
\begin{eqnarray}
A_{x}''+\biggl(\frac{f'}{f}+\frac{b\Phi'\Phi''}{1-b\Phi'^2}\biggr)A_{x}'
+\biggl[\frac{\omega^{2}}{f^{2}}-\frac{k^{2}}{r^{2}f}-\frac{2
\sqrt{1-b\Phi'^2}\textrm{G}(\Psi)}{f}\biggr]A_{x}=0. \label{AxEM}
\end{eqnarray}
We can numerically solve this equation with the appropriate boundary
conditions, i.e., the ingoing wave boundary condition at the event
horizon
\begin{eqnarray}
A_{x}\propto f^{-\frac{i\omega}{3r_{h}}},
\end{eqnarray}
and the asymptotic behavior at the asymptotic AdS boundary
\begin{eqnarray}
A_{x} = A_{x}^{(0)} + \frac{A_{x}^{(1)}}{r} +\cdots.
\end{eqnarray}
Thus, according to the AdS/CFT correspondence, the retarded
correlator has the form \cite{HorowitzPRD78}
\begin{eqnarray}
G_{T}\left(\omega,k \right)=\frac{A_{x}^{(1)}}{A_{x}^{(0)}}.
\end{eqnarray}

In order to calculate $G_{T}^{0}(\omega)$ and $G_{T}^{2}(\omega)$,
we will expand $A_x(r)$ in powers of $k$ just as in Eq. (\ref{GT})
for $G_{T}\left( \omega,k \right)$
\begin{eqnarray}
A_{x}(r)=A_{x0}(r)+k^{2}A_{x2}(r)+\cdots, \label{Expanding}
\end{eqnarray}
which leads to the equations of motion
\begin{eqnarray}
&&A_{x0}''+\biggl(\frac{f'}{f}+\frac{b\Phi'\Phi''}{1-b\Phi'^2}\biggr)A_{x0}'+
\biggl[\frac{\omega^{2}}{f^{2}}-\frac{2
\sqrt{1-b\Phi'^2}\textrm{G}(\Psi)}{f}\biggr]A_{x0}=0, \label{Ax0} \\
&&A_{x2}''+\biggl(\frac{f'}{f}+\frac{b\Phi'\Phi''}{1-b\Phi'^2}\biggr)A_{x2}'
+\biggl[\frac{\omega^{2}}{f^{2}}-\frac{2
\sqrt{1-b\Phi'^2}\textrm{G}(\Psi)}{f}\biggr]A_{x2}-\frac{A_{x0}}{r^{2}f}=0.
\label{Ax2}
\end{eqnarray}
With the asymptotic forms of $A_{x0}$ and $A_{x2}$ found from Eqs.
(\ref{Ax0}) and (\ref{Ax2}), i.e.,
\begin{equation}
A_{x0} = A_{x0}^{(0)} + \frac{A_{x0}^{(1)}}{r} +\cdots, ~~~~~A_{x2}
= A_{x2}^{(0)} + \frac{A_{x2}^{(1)}}{r} +\cdots,
\end{equation}
we get
\begin{equation}
G_{T}^0 (\omega)=\frac{A_{x0}^{(1)}}{A_{x0}^{(0)}}, \ \ \ \ G_{T}^2
(\omega)=\frac{A_{x0}^{(1)}}{A_{x0}^{(0)}}
\biggl[\frac{A_{x2}^{(1)}}{A_{x0}^{(1)}} -
\frac{A_{x2}^{(0)}}{A_{x0}^{(0)}} \biggr]\label{GT0GT2}.
\end{equation}
Substituting Eq. (\ref{GT0GT2}) into Eqs. (\ref{epsmu}) and
(\ref{DLIndex}), we can obtain $\epsilon \left( \omega \right)$,
$\mu\left( \omega \right)$ and $n _{DL}$ in terms of $A_{x0}^{(0)}$,
$A_{x0}^{(1)}$, $A_{x2}^{(0)}$ and $A_{x2}^{(1)}$.

\subsection{Numerical results and discussion}

Using the shooting method, we can solve the equations of motion
(\ref{Ax0}) and (\ref{Ax2}) numerically and then examine the effect
of the Born-Infeld electrodynamics on the negative refraction. For
concreteness, we will set $\xi=0.2$ for the fixed mass of the scalar
field $m^{2}=-2$. It should be noted that  other choices of the
model parameter $\xi$ will not qualitatively modify our results. On
the other hand, we will investigate our system below the critical
temperature, i.e., $T<T_{c}$ since we concentrate on the
superconducting phase.

\begin{figure}[ht]
\includegraphics[scale=0.6]{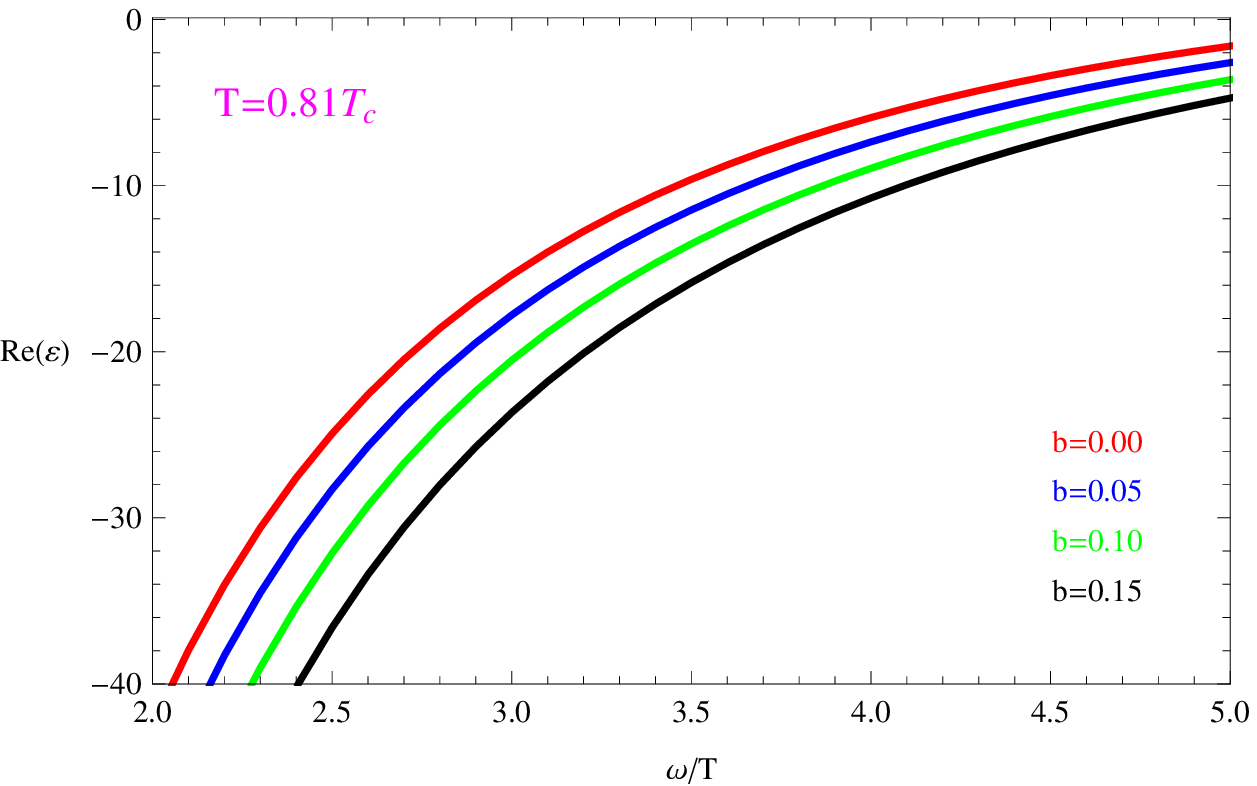}\hspace{0.2cm}%
\includegraphics[scale=0.6]{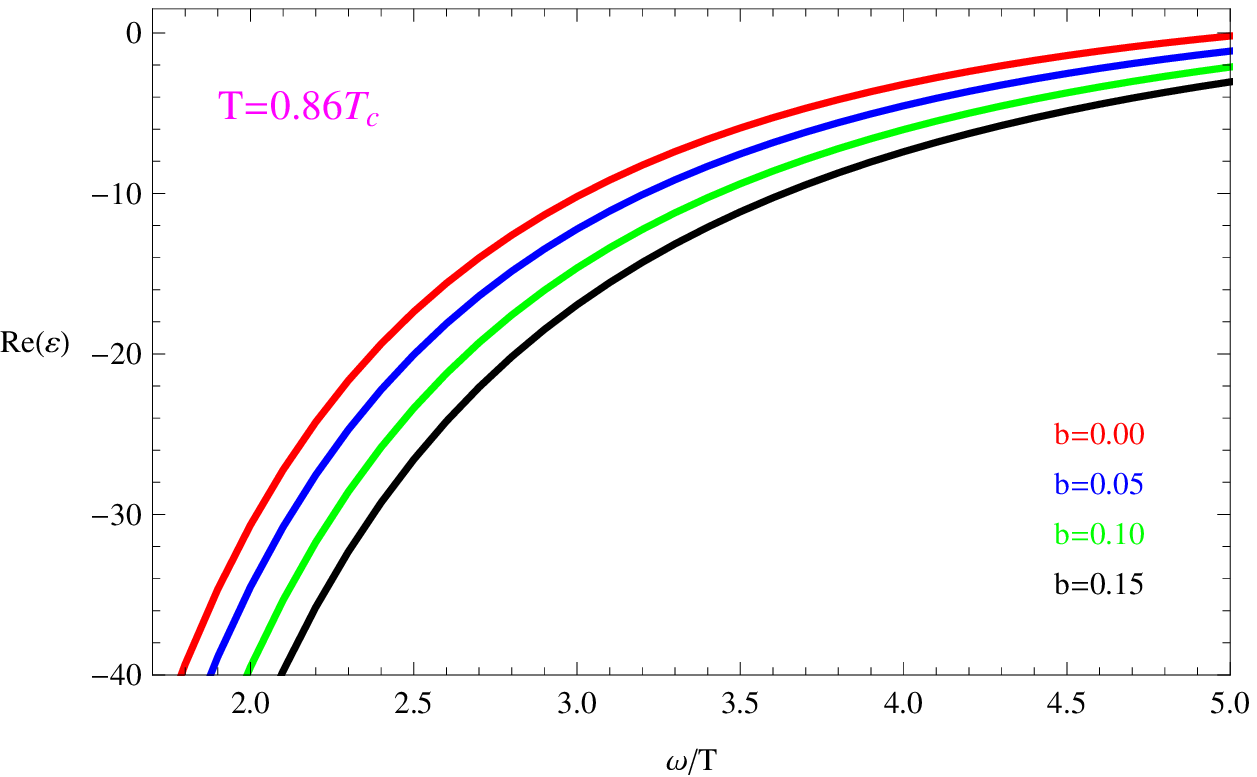}\\ \vspace{0.0cm}
\includegraphics[scale=0.6]{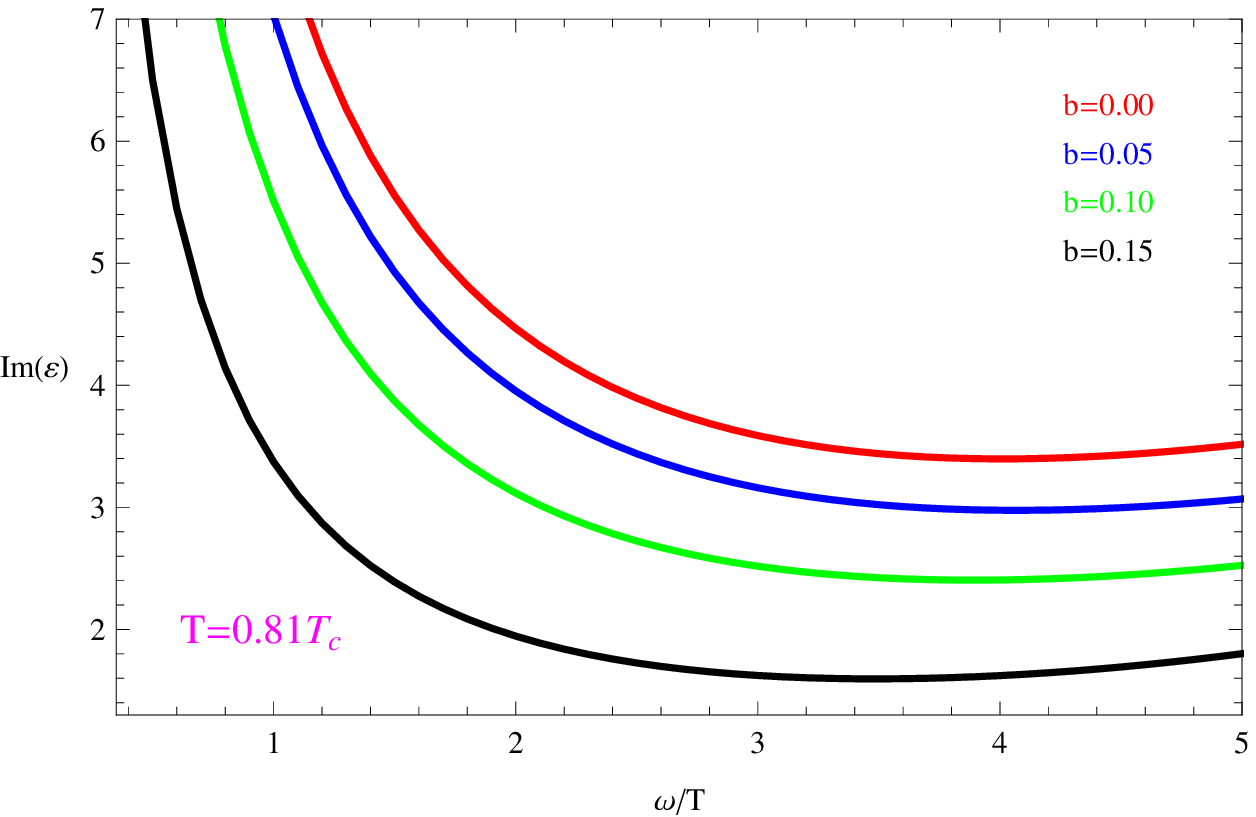}\hspace{0.2cm}%
\includegraphics[scale=0.6]{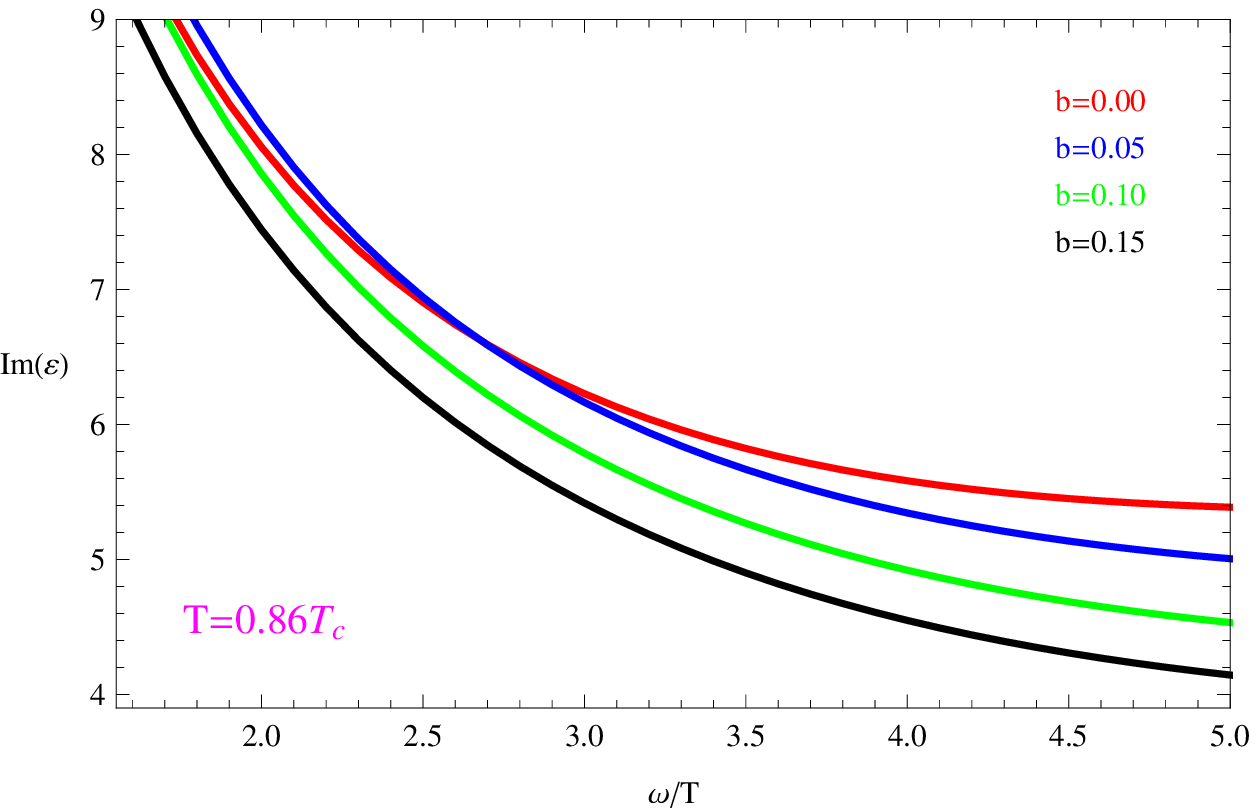}\\ \vspace{0.0cm}
\caption{\label{Permittivity} (Color online) The permittivity
$\epsilon$ as a function of $\omega/T$ with the fixed temperatures
$T=0.81T_{c}$ (left) and $T=0.86T_{c}$ (right) for different values
of the Born-Infeld parameter $b$. The four lines in each panel from
top to bottom correspond to increasing Born-Infeld parameter, i.e.,
$b=0.00$ (red), $0.05$ (blue), $0.10$ (green) and $0.15$ (black)
respectively.}
\end{figure}

\begin{figure}[ht]
\includegraphics[scale=0.6]{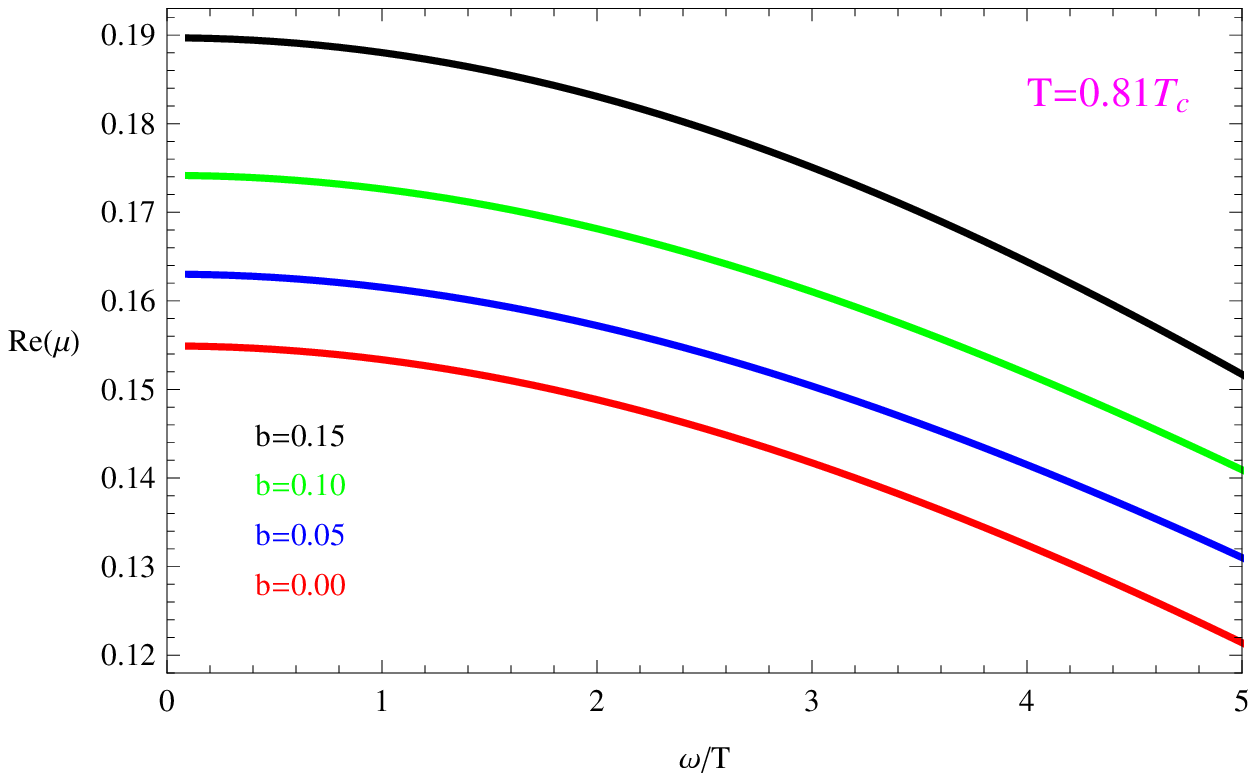}\hspace{0.2cm}%
\includegraphics[scale=0.6]{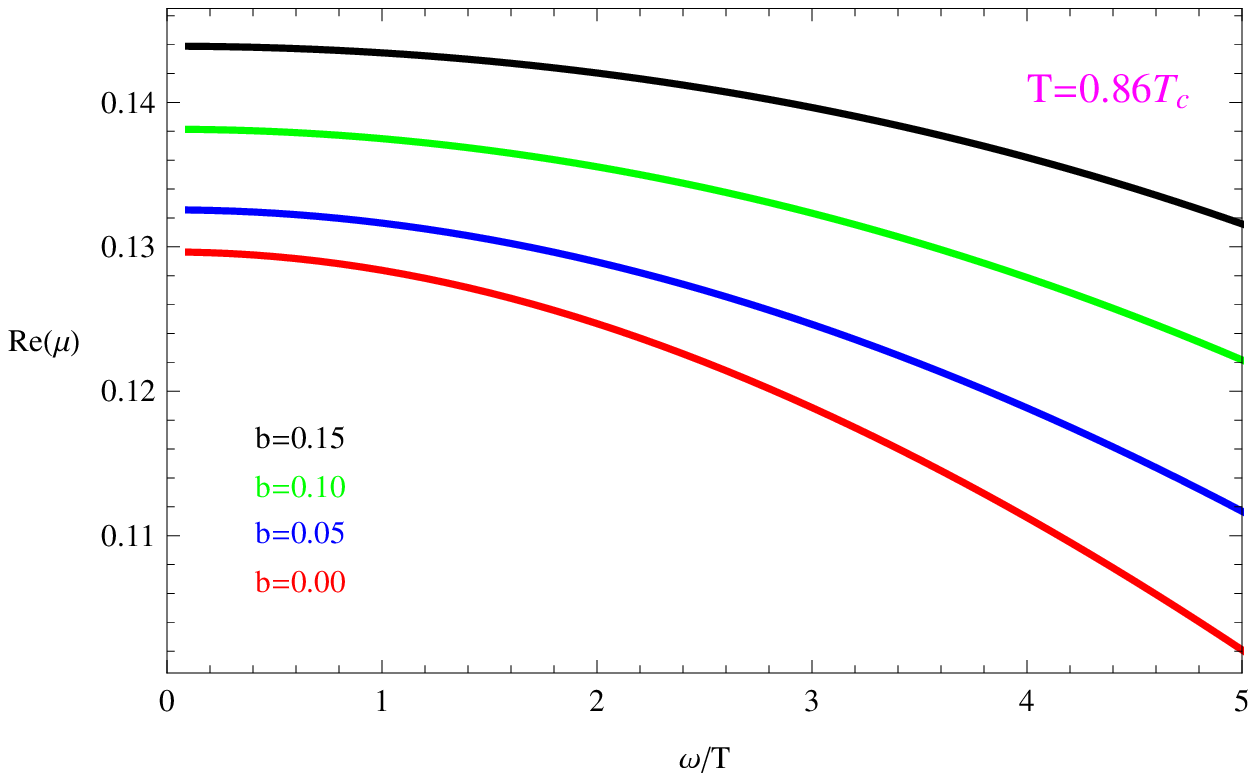}\\ \vspace{0.0cm}
\includegraphics[scale=0.6]{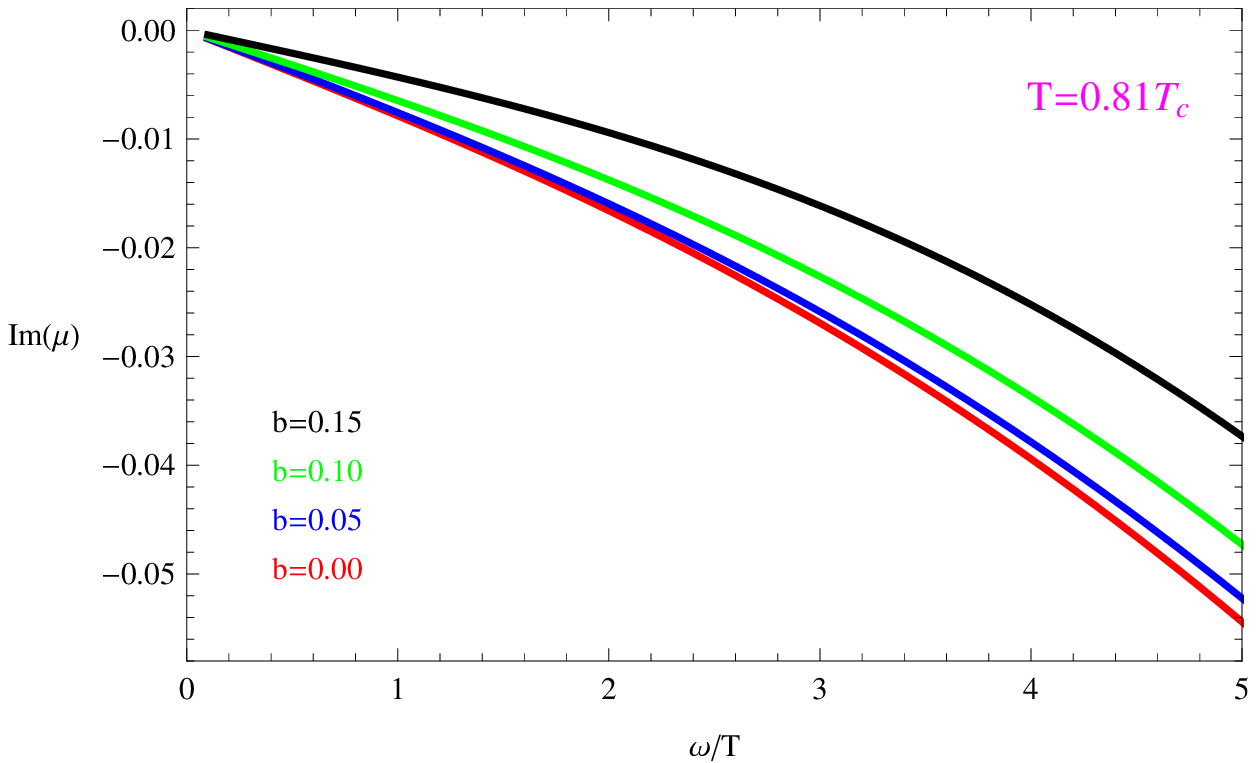}\hspace{0.2cm}%
\includegraphics[scale=0.6]{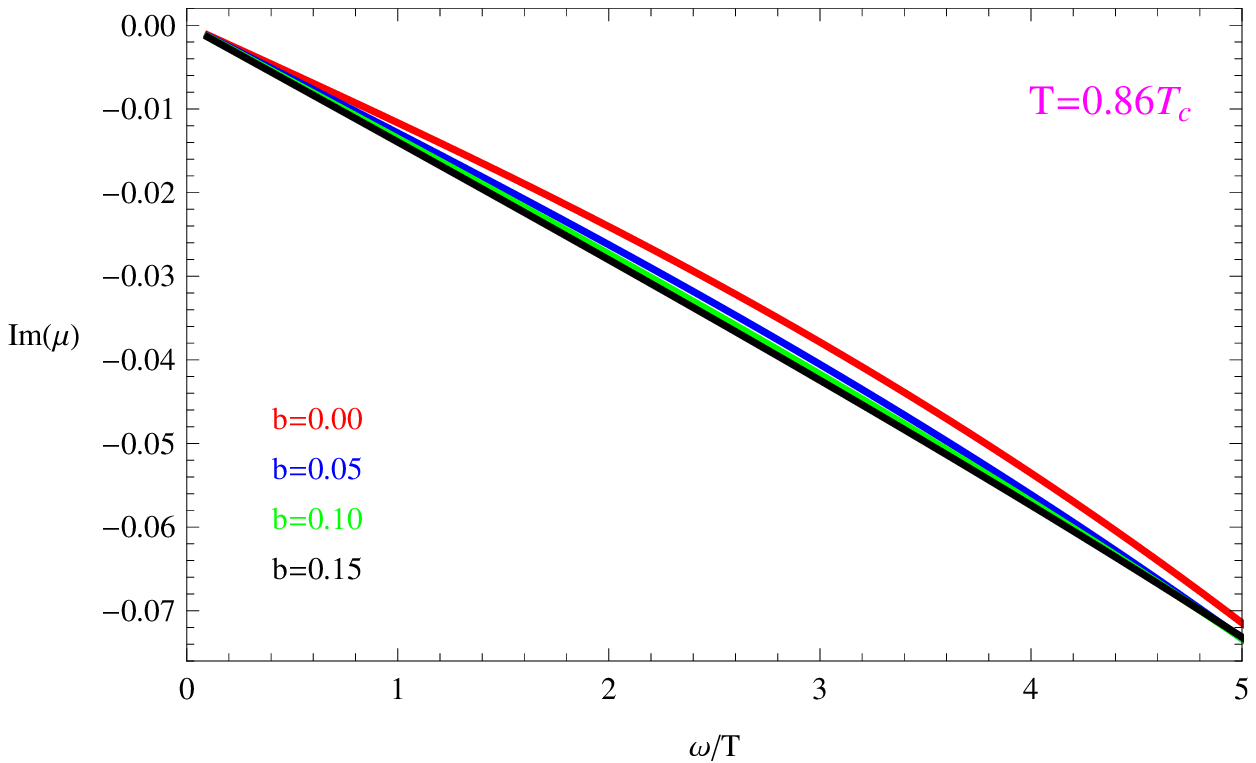}\\ \vspace{0.0cm}
\caption{\label{Permeability} (Color online) The permeability $\mu$
as a function of $\omega/T$ with the fixed temperatures
$T=0.81T_{c}$ (left) and $T=0.86T_{c}$ (right) for different values
of the Born-Infeld parameter $b$. The four lines in each panel
correspond to the Born-Infeld parameter $b=0.00$ (red), $0.05$
(blue), $0.10$ (green) and $0.15$ (black) respectively.}
\end{figure}

In Figs. \ref{Permittivity} and \ref{Permeability}, we plot the
permittivity $\epsilon$ and permeability $\mu$ as a function of
$\omega/T$ with the fixed temperatures $T=0.81T_{c}$ (left) and
$T=0.86T_{c}$ (right) for different values of the Born-Infeld
parameter $b$, i.e., $b=0.00$ (red), $0.05$ (blue), $0.10$ (green)
and $0.15$ (black) respectively. Regardless of the fixed temperature
and Born-Infeld parameter, we observe that, for the permittivity,
Re$(\epsilon)$ is negative at low frequencies and Im$(\epsilon)$ is
always positive and has a pole at the zero frequency, but for the
permeability,  Re$(\mu)$ is always positive and Im$(\mu)$ is negative.
Obviously, Re$(\epsilon)$ and Re$(\mu)$ are not simultaneously
negative, which may be a signal of the negative refraction
\cite{AmaritiFMP2011}. However, just as pointed out in
\cite{AmaritiFMS2011,MahapatraJHEP2014,DeyMT2014}, Im$(\mu)<0$ could
imply some problem in the $\epsilon-\mu$ approach, although we have
defined an effective magnetic permeability that is not an
observable. We will not comment on this further since this delicate
issue is beyond the scope and the purpose of our work.

\begin{figure}[ht]
\includegraphics[scale=0.6]{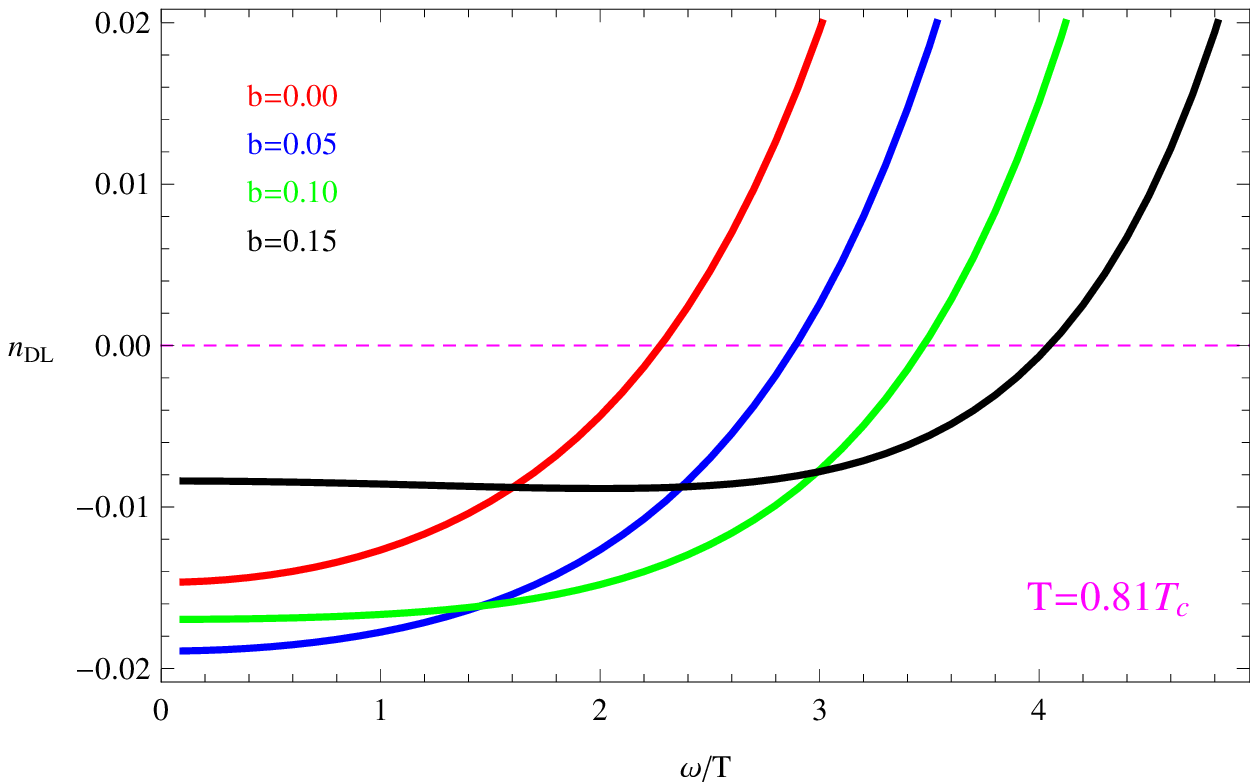}\hspace{0.2cm}%
\includegraphics[scale=0.6]{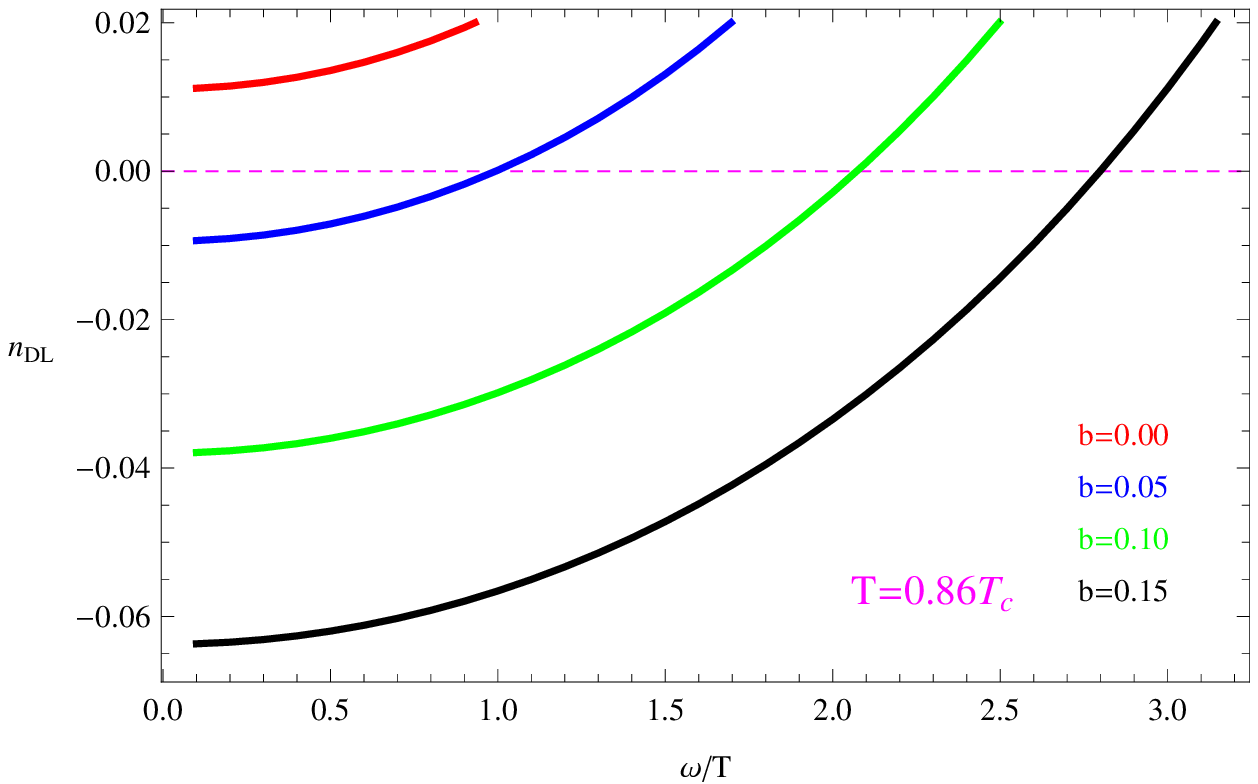}\\ \vspace{0.0cm}
\caption{\label{DLindex} (Color online) The Depine-Lakhtakia index
$n_{DL}$ as a function of $\omega/T$ with the fixed temperatures
$T=0.81T_{c}$ (left) and $T=0.86T_{c}$ (right) for different values
of the Born-Infeld parameter $b$. The four lines in each panel from
top to bottom correspond to increasing Born-Infeld parameter, i.e.,
$b=0.00$ (red), $0.05$ (blue), $0.10$ (green) and $0.15$ (black)
respectively.}
\end{figure}

Our main purpose here is to calculate the DL index $n_{DL}$ and its
dependence on the Born-Infeld parameter $b$. In Fig. \ref{DLindex},
we present $n_{DL}$ as a function of $\omega/T$ with the fixed
temperatures $T=0.81T_{c}$ (left) and $T=0.86T_{c}$ (right) for
different values of the Born-Infeld parameter $b$, i.e., $b=0.00$
(red), $0.05$ (blue), $0.10$ (green) and $0.15$ (black)
respectively. From this Figure, we can see the emergence of  the negative
DL index, below a certain value of $\omega/T$ except in the case of
$b=0$ with $T=0.86T_{c}$. Fixing the temperature of the system, we
find that the  value of $\omega/T$, below which the negative
$n_{DL}$ appears, increases as the Born-Infeld parameter increases,
which indicates that the geater the  Born-Infeld corrections,  the larger  the
range of frequencies for which negative refraction occurs. On the other hand, we note that the negative $n_{DL}$
appears in the case of the Born-Infeld parameter $b=0$ at the
temperature $T=0.81T_{c}$ and disappears at $T=0.86T_{c}$, but the
negative $n_{DL}$ always appears in the case of the Born-Infeld
parameter $b=0.05$ ($b=0.10$ or $b=0.15$) at the temperatures
$T=0.81T_{c}$ and $T=0.86T_{c}$, which shows that  greater
Born-Infeld corrections also make the range of temperatures larger
for which negative refraction occurs. Interestingly, the
Born-Infeld electrodynamics can play an important role in
determining the appearance of negative refraction.

\begin{figure}[ht]
\includegraphics[scale=0.6]{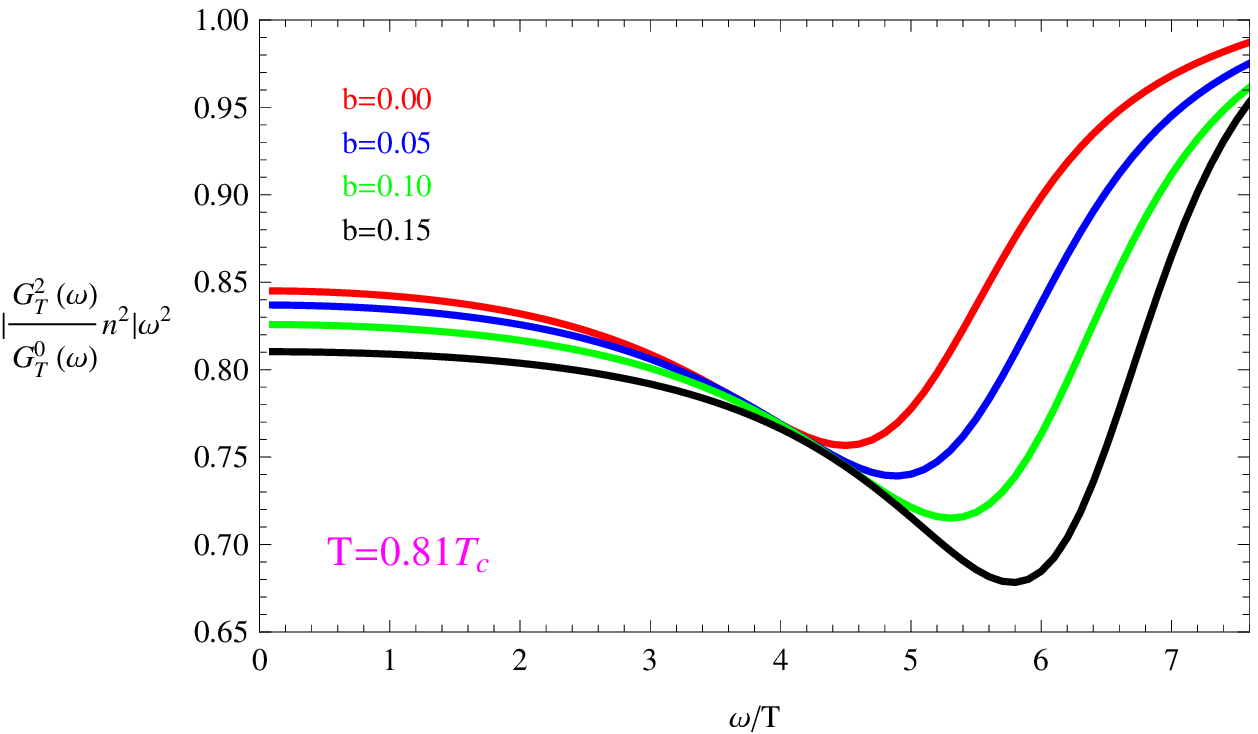}\hspace{0.2cm}%
\includegraphics[scale=0.6]{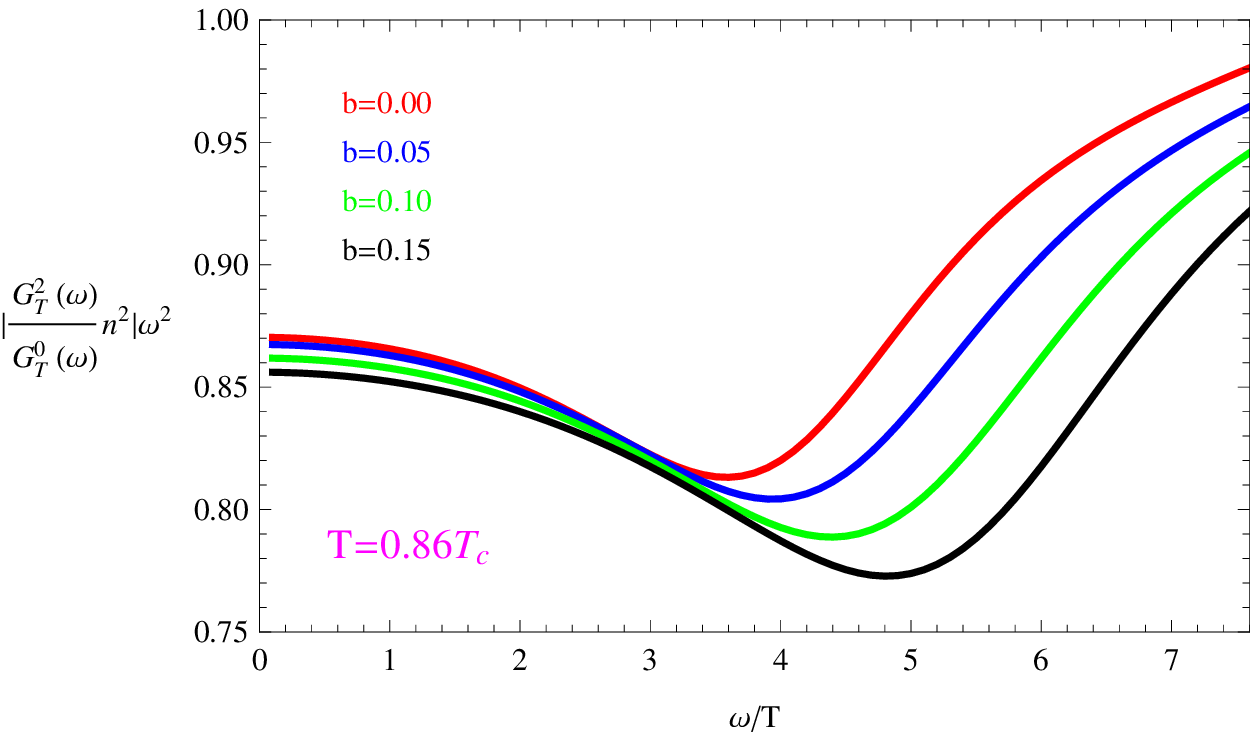}\\ \vspace{0.0cm}
\caption{\label{Constraint} (Color online)
$|\frac{G^{2}_{T}(\omega)}{G^{0}_{T}(\omega)}n^{2}|\omega^{2}$ as a
function of $\omega/T$ with the fixed temperatures $T=0.81T_{c}$
(left) and $T=0.86T_{c}$ (right) for different values of the
Born-Infeld parameter $b$. The four lines in each panel from top to
bottom correspond to increasing Born-Infeld parameter, i.e.,
$b=0.00$ (red), $0.05$ (blue), $0.10$ (green) and $0.15$ (black)
respectively.}
\end{figure}

In order to ensure the validity of the expansion used in Eq.
(\ref{GT}), we require a constraint
\begin{eqnarray}
|\frac{k^{2}G^{2}_{T}(\omega)}{G^{0}_{T}(\omega)}|=|\frac{G^{2}_{T}(\omega)}{G^{0}_{T}(\omega)}n^{2}|\omega^{2}\ll
1. \label{ConstraintEQ}
\end{eqnarray}
Thus, our $\epsilon-\mu$ analysis is valid only for the frequencies
for which this constraint is not violated. The behaviors of
$|\frac{G^{2}_{T}(\omega)}{G^{0}_{T}(\omega)}n^{2}|\omega^{2}$ as a
function of $\omega/T$ with the fixed temperatures $T=0.81T_{c}$
(left) and $T=0.86T_{c}$ (right) for different values of the
Born-Infeld parameter $b$ are given in Fig. \ref{Constraint}, which
shows that, within the plotted frequency range, the constraint
(\ref{ConstraintEQ}) is marginally satisfied in the frequency region
where $n_{DL}$ is negative for all values of $b$ considered here. Of
course, Eq. (23) is not very strictly satisfied and the caveat may
be related to the appearance of a negative imaginary part of the
magnetic permeability, just as pointed out in
\cite{MahapatraJHEP2014,DeyMT2014}. Though Markel proposed that
Im$(\mu)$ (the loss term in the permeability) can in fact be
negative for diamagnetic materials \cite{Markel}, it is worthwhile
to have a better understanding of this point in our holographic
approach. Interestingly, the inclusion of the backreaction can make
Im$(\mu)$ positive
\cite{AmaritiFMS2011,DeyMT2014,MahapatraJHEP2015}. We will leave
this subject for future research.

\begin{figure}[ht]
\includegraphics[scale=0.6]{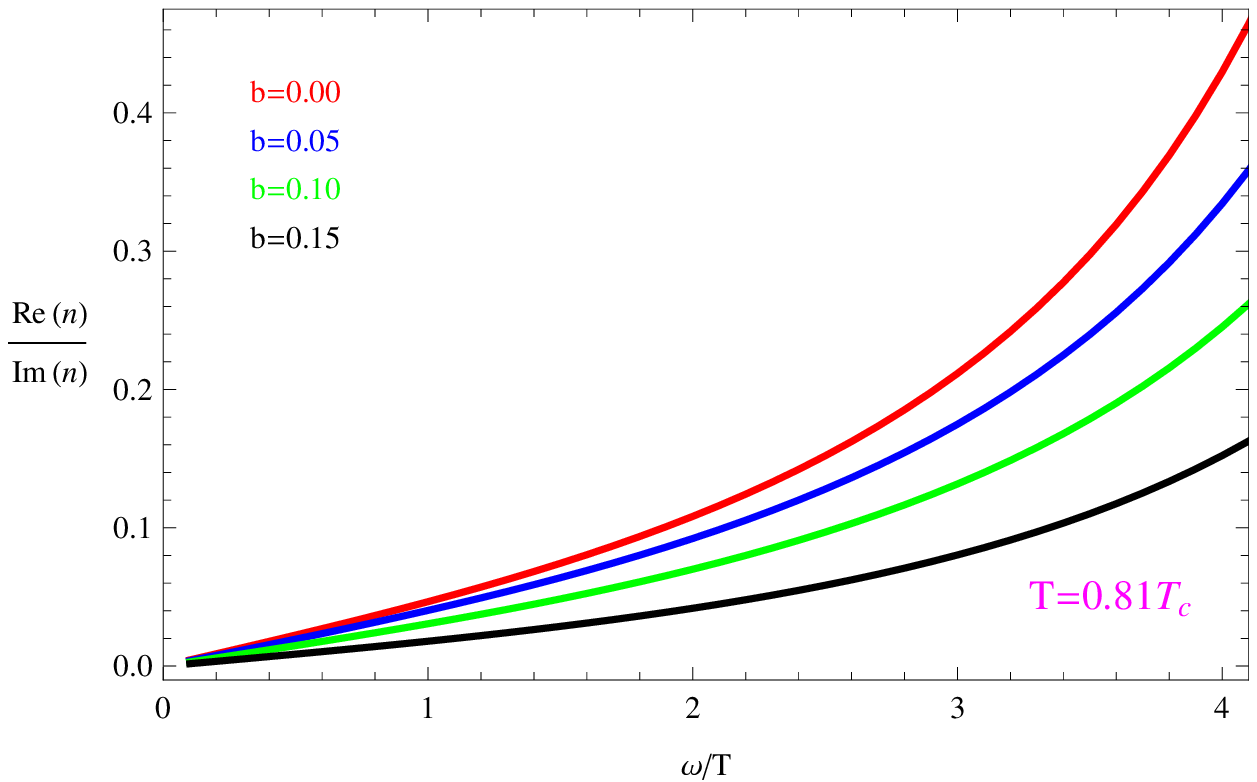}\hspace{0.2cm}%
\includegraphics[scale=0.6]{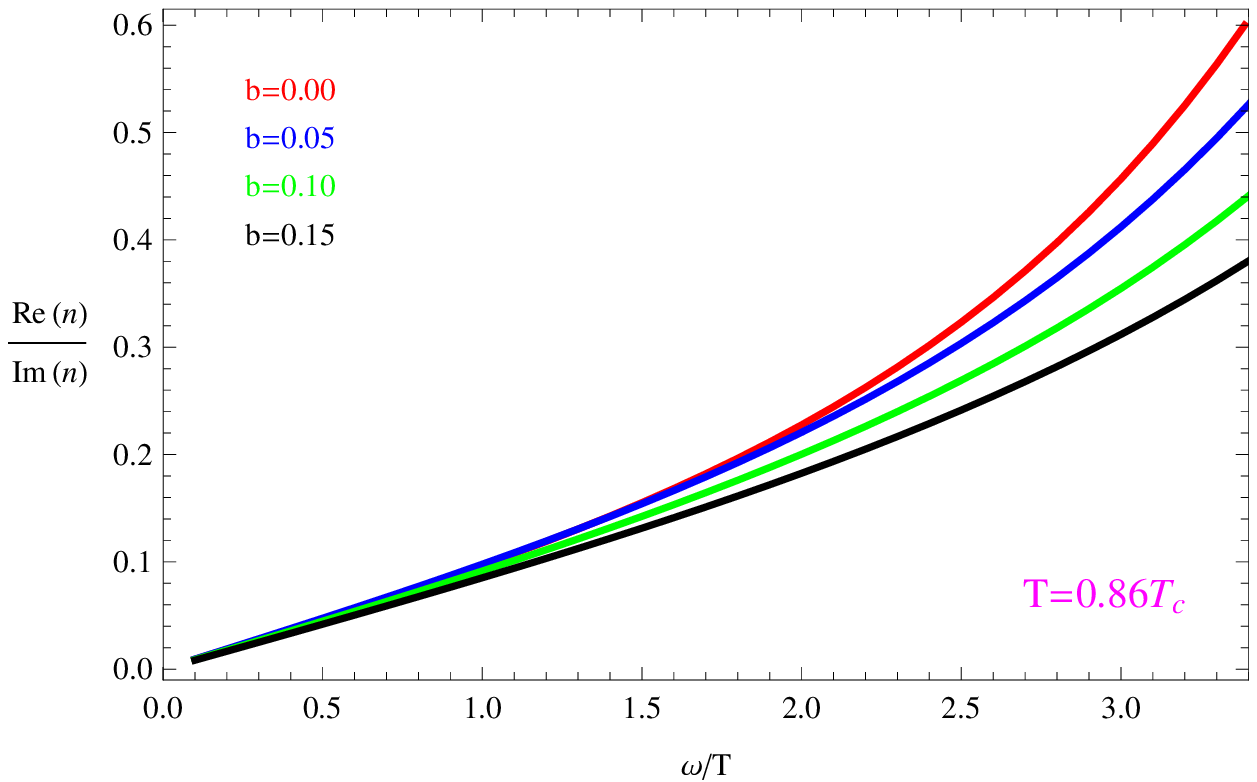}\\ \vspace{0.0cm}
\caption{\label{Dissipation} (Color online) The ratio $Re(n)/Im(n)$
as a function of $\omega/T$ with the fixed temperatures
$T=0.81T_{c}$ (left) and $T=0.86T_{c}$ (right) for different values
of the Born-Infeld parameter $b$. The four lines in each panel from
top to bottom correspond to increasing Born-Infeld parameter, i.e.,
$b=0.00$ (red), $0.05$ (blue), $0.10$ (green) and $0.15$ (black)
respectively.}
\end{figure}

Finally, we consider the ratio $Re(n)/Im(n)$ and discuss the
dissipation effects in our system. In Fig. \ref{Dissipation}, we
plot the ratio $Re(n)/Im(n)$ as a function of $\omega/T$ with the
fixed temperatures $T=0.81T_{c}$ (left) and $T=0.86T_{c}$ (right)
for different values of the Born-Infeld parameter $b$, i.e.,
$b=0.00$ (red), $0.05$ (blue), $0.10$ (green) and $0.15$ (black)
respectively. For the fixed temperature, we observe that the ratio
decreases with increasing values of $b$ for the fixed $\omega/T$, and
the magnitude of $Re(n)/Im(n)$ is small within the negative
refraction frequency range, which implies large dissipation in the
system. However, we can use the  Born-Infeld corrections to
reduce the dissipation since the negative refraction
frequency range  depends on the Born-Infeld parameter $b$. For
example, the ratio $Re(n)/Im(n)$ is about $0.10$ when $\omega/T=2.0$
with $b=0.00$ but is about $0.13$ when $\omega/T=2.6$ with $b=0.05$
for the case of $T=0.81T_{c}$. Obviously, the Born-Infeld
electrodynamics can be used to improve the propagation in the
holographic setup.

\section{conclusions}

We have constructed the generalized superconductors with the Born-Infeld
electrodynamics and studied their negative refraction in the
probe limit, which may help us to understand the influences of the
$1/N$ or $1/\lambda$ corrections on the holographic superconductor
models and their optical properties. Varying the Born-Infeld
parameter as well as the temperature, we calculated in details the
electric permittivity, the effective magnetic permeability, the refractive
index, and the Depine-Lakhtakia (DL) index of our system and observed
the existence of negative refraction in the superconducting phase at
small frequencies. Interestingly, we found that the greater
the Born-Infeld corrections the larger the range of frequencies or the range
of temperatures  for which a negative DL index occurs,
which indicates that the Born-Infeld electrodynamics facilitates the appearance of negative refraction.
Furthermore, we analyzed the dissipation effects in our system and found that the tunable Born-Infeld parameter can be used to improve
the propagation in the holographic setup. Thus, we concluded that
the Born-Infeld electrodynamics can play an important role in
determining the optical properties of the boundary theory. The
extension of this work to the fully backreacted spacetime would be
interesting since the backreaction provides richer physics in the
generalized holographic superconductors and significantly
affects their optical properties \cite{DeyMT2014,MahapatraJHEP2015}. We
will leave this for future study.

{\bf Note added}------While we were completing this work, a
complementary paper \cite{Zangeneh} on the optical properties of
Born-Infeld-dilaton-Lifshitz holographic superconductors appeared in
arXiv.

\begin{acknowledgments}

This work was supported by the National Natural Science Foundation
of China under Grant Nos. 11775076, 11690034 and 11475061; Hunan
Provincial Natural Science Foundation of China under Grant No.
2016JJ1012.

\end{acknowledgments}

\end{document}